\documentclass[12pt, amsmath, amssymb, nofootinbib, sort&compress, showkeys, round, citeautoscript]{revtex4}

\setcitestyle{square}

\usepackage{graphics}
\usepackage{epsfig}
\usepackage{subfigure}  
\usepackage{bm}
\usepackage{booktabs}       
\usepackage{threeparttable} 

\makeatletter
\def\@biblabel#1{(#1)}
\makeatother

\begin{document}
\title{ZnSnS$_3$: Structure Prediction, Ferroelectricity, and Solar Cell Applications}

\author{
Radi A. Jishi}

\email{rjishi@calstatela.edu}

\affiliation{ Department of Physics, California State University, Los Angeles, California, U.S.A}

\author{Marcus A. Lucas}

\email{mlucas9@calstatela.edu}

\affiliation{ Department of Electrical Engineering, California State University, Los Angeles, California, U.S.A}


\date{\today}

\begin{abstract}

The rapid growth of the solar energy industry has produced a strong demand for high performance, 
efficient photoelectric materials. Many ferroelectrics, composed of earth-abundant
 elements,  are useful for solar cell applications due to their large internal polarization. 
However, their wide band gaps prevent them from absorbing light in the visible to mid-infrared 
range. Here, we address the band gap issue by investigating, in particular, the substitution of 
sulphur for oxygen in the perovskite structure ZnSnO$_3$. Using evolutionary methods we identify 
the stable and metastable structures of ZnSnS$_3$ and compare them to those previously characterized 
for ZnSnO$_3$. Our results suggest that ZnSnS$_3$ forms a monoclinic structure followed by 
metastable ilmenite and lithium-niobate  structures. The latter structure is highly polarized and it
possesses a significantly reduced band gap of 1.28 eV. These desirable characteristics make it a prime
candidate for solar cell applications.

\end{abstract}


\maketitle

\section{Introduction}\label{sec:1}

Ferroelectrics are materials that possess spontaneous electric polarization. This results
from a lack of inversion symmetry; all ferroelectric crystals are non-centrosymmetric.
Due to intrinsic polarization, ferroelectrics may serve as light harvesters in photovoltaic
devices \cite{ref:Cao, ref:Alexe, ref:Qin, ref:Choi1, ref:Butler, ref:Yuan1}. In a
semiconductor p-n junction acting as a photovoltaic device, the built-in potential
across the depletion layer is used to separate the photoexcited electron-hole pairs; the
maximum open-circuit voltage is thus almost equal to the semiconductor band gap. In a
ferroelectric, on the other hand, the separation of the photoexcited pairs is due
to the built-in potential induced by the intrinsic polarization; this makes possible
the realization of open-circuit voltages that far exceed the band gap \cite{ref:Yang,
ref:Inoue, ref:Young, ref:Huang, ref:Yuan2, ref:Bennett, ref:Fridkin1,
ref:Fridkin2, ref:Glass, ref:Chynoweth, ref:Neumark}.

Among the most important ferroelectrics are metal oxide perovskites with
the general formula ABO$_3$, where A and B are metal cations (usually, B
is a transition metal). Well-known examples of such ferroelectrics are
BaTiO$_3$ and LiNbO$_3$. These oxides have relatively large internal 
electric fields that could be exploited in photovoltaic applications. However,
progress in this area has been hampered by the fact that these ferroelectrics have a
large band gap (3-4 eV), which makes them unsuitable to function as  efficient
light harvesters. The large band gap is due to the strong ionic bonding between
the transition metal B and oxygen, which, in turn, is due to the large
difference in electronegativity between these atoms. In ABO$_3$, the highest
valence band is derived from oxygen 2p orbitals, while the low conduction
bands are derived from the transition metal 3d states.

To reduce the band gap in ferroelectrics, different strategies have been implemented.
Choi et al. \cite{ref:Choi2} alloyed the ferroelectric Bi$_4$Ti$_3$O$_{12}$, which
has an optical band gap between 3.1 and 3.6 eV  \cite{ref:Ehara, ref:Singh, ref:Jia},
with LaCoO$_3$, which is a Mott insulator with a small band gap of 0.1 eV \cite{ref:Arima}.
From X-ray diffraction and scanning transmission electron microscopy they concluded
that some La atoms substitute for some Bi atoms at some specific sites, and some Co atoms
occupy some Ti sites. Since Co is more electronegative than Ti (1.88 vs 1.54),
Co-O bond is less ionic than Ti-O bond, and a reduction in the band gap is expected. Indeed,
a reduction of up to 1 eV was observed.

In another approach, Grinberg et al. \cite{ref:Grinberg} placed two
different transition metal cations on the perovskite B-site, with
one atom driving ferroelectricity and the other producing a band gap in
the visible range. They mixed the ferroelectric oxide potassium niobate (KNbO$_3$)
with barium nickel niobate (BaNi$_{1/2}$Nb$_{1/2}$O$_{3-\delta}$) so as to introduce
Ni$^{2+}$ on the B-site along with an oxygen vacancy which can give rise to gap states
in the host KNbO$_3$ crystal. The solid solutions thus formed,
[KNbO$_3$]$_{1-x}$[BaNi$_{1/2}$Nb$_{1/2}$O$_{3-\delta}$]$_x$ with $x=0.1$ to $0.5$, were
ferroelectric with a direct band gap ranging from 1.1-3.38 eV.

In some multiferroics, which exhibit a magnetic order along side the ferroelectric one, a
somewhat smaller band gap exists. In BiFeO$_3$ the band gap is
2.7 eV \cite{ref:Basu, ref:Hauser}. It is thus expected that multiferroics, such as
BiFeO$_3$ and Bi$_2$FeCrO$_6$ \cite{ref:Nechache1, ref:Nechache2}, would be
more promising candidates for solar cell applications. In  Bi$_2$FeCrO$_6$
epitaxial thin films, the optical band gap depends on the degree of Fe-Cr
ordering. This dependence results from the hybridization of the 3d orbitals
in Fe and Cr with the 2p orbitals in oxygen. Nechache et al.  \cite{ref:Nechache3}
investigated the effect on the band gap of Fe/Cr ordering and found that, under
the right film growth conditions, the band gap can be tuned all the way from 2.7 eV
down to 1.5 eV.

An important class of ferroelectric oxides are those that crystallize in the
LiNbO$_3$ structure (LN-structure) with space group R3c. These crystals are
non-centrosymmetric with a large polarization. LN-ZnSnO$_3$, synthesized
under a pressure of 7 GPa \cite{ref:Inaguma1}, has a polarization given by
59 $\mu$C/cm$^2$. Other LN-type polar oxides synthesized under high pressure
include CdPbO$_3$ \cite{ref:Inaguma2}, PbNiO$_3$ \cite{ref:Inaguma2, ref:Inaguma3},
GaFeO$_3$ \cite{ref:Arielly}, and LiOsO$_3$ \cite{ref:Shi}. Ilmenite ZnGeO$_3$
transforms to an orthorhombic perovskite phase at 30 GPa, and upon releasing
the pressure, to the LN-structure \cite{ref:Yusa, ref:Akaogi1} with a polarization
of about 60 $\mu$C/cm$^2$  \cite{ref:Zhang, ref:Inaguma4}. Several other LN-type
oxides, such as MnTiO$_3$ \cite{ref:Syono1, ref:Ito, ref:Ko, ref:Ross},
MnSnO$_3$ \cite{ref:Syono2, ref:Leinenweber1},
FeTiO$_3$ \cite{ref:Ito, ref:Leinenweber1, ref:Mehta, ref:Leinenweber2},
FeGeO$_3$ \cite{ref:Hattori}, 
MgGeO$_3$ \cite{ref:Akaogi1, ref:Ito, ref:Leinenweber3}, and CuTaO$_3 $ \cite{ref:Sleight}
were similarly obtained during decompression from the perovskite-type phase, which
is stable at high pressure. More recently, LN-type ZnTiO$_3$, with a large
polarization of 88 $\mu$C/cm$^2$, has been synthesized in this fashion \cite{ref:Akaogi2}.

Polar oxide semiconductors of the LN-type have wide band gaps. For example, ZnSnO$_3$ has a band gap
of 3.3-3.7 eV \cite{ref:Borhade, ref:Mizoguchi}; such values are typical for this class
of materials. However, it has been noted that the band gap in ZnSnO$_3$ is very
sensitive to variations in the lattice constants, and suggestions have been made to
tune the band gap by growing ZnSnO$_3$ films on a substrate with some degree of
lattice mismatch \cite{ref:Kolb1}, by substituting sulfur for oxygen \cite{ref:Kolb2},
or by substitutional doping with calcium or barium \cite{ref:Kons}.

In this work, we carry out first-principles calculations,
using density functional theory (DFT), on three LN-type
crystals with large remnant polarization, namely, ZnSnO$_3$, ZnGeO$_3$,
and ZnTiO$_3$. We show
that, upon using the modified Becke-Johnson (mBJ) exchange potential \cite{ref:Tran09},
the correct sizes of the band gaps are calculated for these crystals. Having established the 
validity of this computational method in producing the correct band gaps for this family of 
compounds, we then focus on ZnSnO$_3$ and study the efect of substituting sulfur for oxygen. 
Since the structure of the resulting compound, ZnSnS$_3$, is unknown, we carry out extensive 
calculations using evolutionary algorithms in order to determine its structure. 
 We find that the most stable structure has a monoclinic unit cell, followed 
by two metastable structures, namely an ilmenite and a lithium niobate (LN) structure. The 
monoclinic and ilmenite phases are not polar, but the LN-phase is ferroelectric with a 
large polarization and a small band gap of 1.28 eV.

\section{Methods}\label{sec:2}

To predict the stable and metastable structures of ZnSnS$_3$, two different evolutionary
methods, implemented in the codes USPEX and CALYPSO, were employed. The USPEX (Universal
Structure Predictor: Evolutionary Xtallography) code 
\cite{ref:Oganov1, ref:Lyakhov, ref:Oganov2}, developed by Oganov, Glass, Lyakhov and Zhu,
features local optimization, real-space representation and variation operators that mimic
natural evolution. The CALYPSO (Crystal structure AnaLYsis by Particle Swarm Optimization)
code  \cite{ref:calypso}, developed by Wang, Lv, Zhu and Ma, uses local structural optimization
and the particle swarm optimization (PSO) method to update structures.

The first step, in both methods, is to generate a population of random crystal structures,
each with a symmetry described by a randomly chosen space group. Once a space group is selected,
appropriate lattice vectors and atomic positions are generated. Each generated structure is optimized
using density functional theory and its free energy (known as the fitness function) is calculated.
The structure optimization is carried out using the code VASP \cite{ref:Kreese1, ref:Kreese2, ref:Kreese3}, which uses a plane wave basis for
expanding the electronic wave function. Each structure is optimized in four steps, beginning with a
coarse optimization and gradually increasing the accuracy. In the last optimization step, the kinetic
energy cutoff for plane wave expansion of the wave function is 600 eV. The optimized structures of
the initially generated population constitute the first generation, each member being called an
individual. A new generation is then produced
where some of its members are generated randomly while others are obtained from the best structures
(those with lowest energy) of the previous generation. In USPEX, new individuals (offspring) are
produced from parent structures by applying variation operators such as heredity, mutation, or
permutation. In the PSO method, a new structure is generated from a previous one by updating
the atomic positions using an evolutionary algorithm. The structures in the new generation are
optimized, and the best among them serve as precursors for structures in the next generation.
The process continues until convergence to the best structures is achieved.

Band gaps, band structures, and densities of states were calculated using the all-electron,
full-potential, linearized, augmented plane wave method as implemented in the WIEN2K code  \cite{ref:Blaha}.
Here, space is divided into two regions. One region comprises the interior of nonoverlapping
muffin-tin spheres centered on the atomic sites, while the rest of the space (the interstitial)
makes the other region. The electronic wave function is expanded in terms of a set of basis
functions which take different forms in the two distinct regions of space. Inside the spheres,
the basis functions are atomic-like functions written as an expansion in spherical harmonics
up to $l_{max} = 10$. In the interstitial, they are plane waves with a maximum wave vector
of magnitude K$_{max}$. Each plane wave is augmented by one atomic-like function in each
muffin-tin sphere. K$_{max}$ was chosen so that R$_{mt}$K$_{max}$ $= 8$, where R$_{mt}$
is the radius of the smallest muffin-tin sphere in the unit cell. Charge density was
Fourier-expanded up to a maximum wave vector of 14 $a_0^{-1}$, where $a_0$ is the Bohr radius.
Modified Becke-Johnson exchange potential  \cite{ref:Tran09}, which is known to yield reasonably accurate
band gap values in semiconductors, was adopted in our band structure calculations.
In the self-consistent field calculations, the total energy and charge were converged to within
0.1 mRy and 0.001 e, respectively. For total energy calculations, a Monkhorst-Pack  \cite{ref:Monkhorst}
8x8x8 grid of \textbf{k}-points in the Brillouin zone is used.

\section{Results and Discussion}\label{sec:3}

As a rough guide to predicting the stable structure of a compound of the form ABX$_3$,
where A and B are cations and X is an anion, one usually calculates the Goldschmidt
tolerance factor given by

\begin{equation}
 t = \frac{r_A + r_X}{\sqrt{2}(r_B + r_X)}
\end{equation}

where $r_A$, $r_B$, and $r_X$ are the radii of the A, B, and X ions, respectively.
For Zn, Sn, O, and S, the ionic radii are given, respectively, by 0.74 \AA, 0.69 \AA,
1.4 \AA, and 1.84 \AA. Using these values, we find that for ZnSnO$_3$, $t = 0.724$,
while for ZnSnS$_3$, $t = 0.721$. We are thus tempted to conclude that the replacement
of oxygen with sulfur should not cause a change in structure. As stated in the introduction,
the most stable phase of ZnSnO$_3$ is the ilmenite structure (space group R-3) which,
upon application and subsequent release of pressure, transforms to the ferroelectric lithium niobate (LN) phase
(space group R3c). One would expect ZnSnS$_3$ to behave in a similar way. However, 
the large polarizability of sulfur indicates that the Goldschmidt
tolerance factor may not be sufficient for predicting the structure
of sulfur-containing compounds  \cite{ref:Brehm}.

To resolve this issue, we carried out extensive calculations using the evolutionary
algorithm implemented in USPEX and the PSO method used by CALYPSO. During the USPEX calculation,
20 generations were produced with each generation containing 20 individuals. In the PSO
method we produced 20 generations, each generation consisting of 50 individuals.
For all generated structures, the unit cell was assumed to contain two formula units. The
 most stable structures were identified, each with a specific space group. To analyze
the results further, 80 random structures were subsequently generated for each of the most
stable space groups. Out of each set of 80 optimized structures, the single structure with 
the lowest energy was identified. In this fashion, the best structures of ZnSnS$_3$, along with
their specific space groups, were determined.

In Table 1 we list these ZnSnS$_3$ structures by order of their stability. To compare the 
relative stability of various structures, the free energy per atom in the most
stable structure is set equal to zero. That structure has a monoclinic unit cell 
with space group P2$_1$ (number 4). The metastable structure with lowest energy is the
ilmenite, followed by the LN-type structure. This is to be contrasted with ZnSnO$_3$, where
the ilmenite structure is most stable and the LN-phase is a metastable one. Similar to the
case of ZnSnO$_3$, we expect that ZnSnS$_3$ would undergo a phase transition from the most 
stable monoclinic phase to the metastable phase of ilmenite or LN-type under appropriate 
conditions of temperature and pressure.

\begin{table}[ht]
  \caption{The predicted most stable structures of ZnSnS$_3$ at atmospheric pressure. The
space groups that describe the symmetry of the structures, the lattice constants (in \AA),
the angles between the lattice vectors (in degrees), and the relative energies per atom
(in eV) are shown. For ease of comparison, the energy per atom of the most stable
structure is set equal to zero.}
    \begin{threeparttable}
      \begin{tabular}{ccc}
	\hline
	\hline
	Space group (number) & Lattice constants (\AA) and angles (degrees) & Energy/atom (eV)\\
	\hline
	 P2$_1$ (4)         &  a=8.714, b=6.398, c=3.752, $\alpha =\beta = 90, \gamma = 92.362$  	& 0.000			\\
         R-3 (148)            & a=b=c=7.016, $\alpha = \beta = \gamma = 53.689$		& 0.040				\\
	 R3c (161)            & a=b=c=6.761, $\alpha = \beta = \gamma = 55.799$	& 0.068 				\\
	 P6$_3$/m (176)    & a=b=3.737, c=17.405, $\alpha = \beta = 90, \gamma = 120$ 	& 0.074	\\
         R32     (155)         & a=b=c=6.828, $\alpha = \beta = \gamma = 55.122$     	& 0.089	\\
         P6$_3$  (173)      & a=b=6.187, c=6.431, $\alpha = \beta = 90, \gamma = 120$	        & 0.100	\\
         Cc (9)        & a=b=6.393, c=10.796, $\alpha = \beta = 88.229, \gamma = 146.963$                 & 0.145       \\
         P6$_3$mc (186)		    & a=b=3.766, c=16.559, $\alpha = \beta = 90, \gamma = 120$		& 0.154				\\
         R-3c (167)    &  a=b=c=6.794, $\alpha = \beta = \gamma = 56.359$                & 0.155        \\
	\hline
	\hline
     \end{tabular}
    \label{tab:1}
   \end{threeparttable}
\end{table}

To evaluate the relevance of ZnSnS$_3$ to solar cell applications, we have calculated the
band structures of the three most stable phases of this compound. Since low band gap values
for the light absorbers are essential for photovoltaic applications, it is important that
the computational method yield accurate values for these gaps. Density functional theory,
when using the local density approximation (LDA) or generalized gradient approximation (GGA)
to approximate the exchange-correlation term, is known to severely underestimate band gaps
in semiconductors. The use of the modified Becke-Johnson  (mBJ) exchange potential results
in a great improvement in the band gap values comparable to that achieved by the much more
expensive GW approximation. For example, consider the case of ZnSnO$_3$. It has a measured
band gap E$_g$ in the range 3.3-3.7 eV  \cite{ref:Borhade, ref:Mizoguchi}.
Our calculations, using GGA, give a band gap of 1.5 eV. On the other hand, using the mBJ 
exchange potential, we calculate a band gap of 3.4 eV, which is in excellent agreement with 
experiment. We also calculated the band gaps in ZnGeO$_3$ and ZnTiO$_3$ using this same method,
and found them to be 3.57 eV and 3.51 eV, respectively. Though experimental values are not 
available for the band gaps of theses two compounds, the calculated values are typical of those 
encountered in ferroelectric oxides.

In Fig. 1 we present the calculated energy bands and density of states of the monoclinic phase
of ZnSnS$_3$, with space group P2$_1$, using the mBJ exchange potential. The band gap is 1.90 eV.
Similar calculations on the ilmenite phase of ZnSnS$_3$, with the space group R-3, feature a
band gap of 2.30 eV. These two phases of ZnSnS$_3$ are not very useful as light harvesters in solar
cells for two reasons: (1) they do not have small band gaps, and (2) they are not polar.

 \begin{figure}[htbp]
  \includegraphics[height=9.5cm]{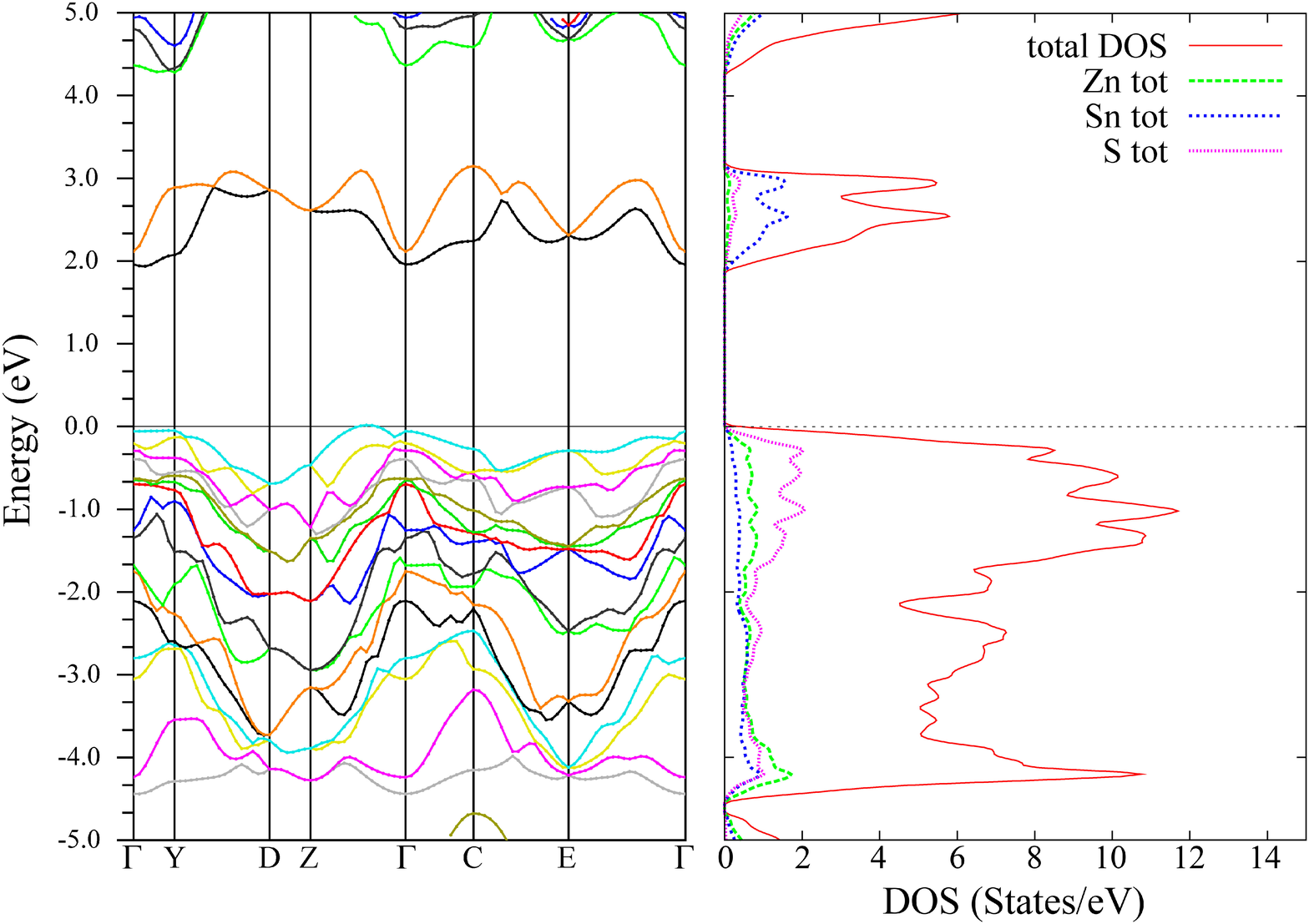}
   \caption{ Energy bands in the monoclinic phase (space group number 4) of ZnSnS$_3$,
     plotted along high symmetry directions, along with a plot of the density of states.
     The bands are calculated by using the mBJ potential for exchange and correlation.
     The zero of energy is taken to coincide with the valence band maximum.
  \label{fig:1}}
 \end{figure}

The situation is markedly different for the LN-phase of ZnSnS$_3$ with  space group R3c (number 161).
This phase has a low band gap and a large polarization. In Fig. 2 we show the calculated
band structure, using the mBJ potential, along with the density of states. Here the band gap is
1.28 eV, in agreement with a previous calculation  \cite{ref:Kolb2} that used the GW approximation.
The upper valence bands are derived mainly from S p-orbitals while the lower conduction bands
result mainly from Sn s- and S p-orbitals. The polarization, calculated using the Berry phase
formalism of the modern theory of crystal polarization  \cite{ref:Resta, ref:King}, is found
to be 57  $\mu$C/cm$^2$, which is essentially the same as in the ferroelectric ZnSnO$_3$.
About one third of the polarization is electronic and the rest is ionic. The large ionic polarization
derives from the particular crystal structure of this phase of ZnSnS$_3$. Though the Sn ion
is octahedrally bonded to six S ions, it does not sit at the center of the octahedron; rather,
three of the six Sn-S bonds have a length of 2.42 \AA while the other three have a length of 2.57 \AA.
Hybridization between the Sn s- and S p-orbitals causes a displacement of the Sn ion from the center 
of the octahedron, thereby lowering the energy of the system.

\begin{figure}[htbp]
  \includegraphics[height=9.5cm]{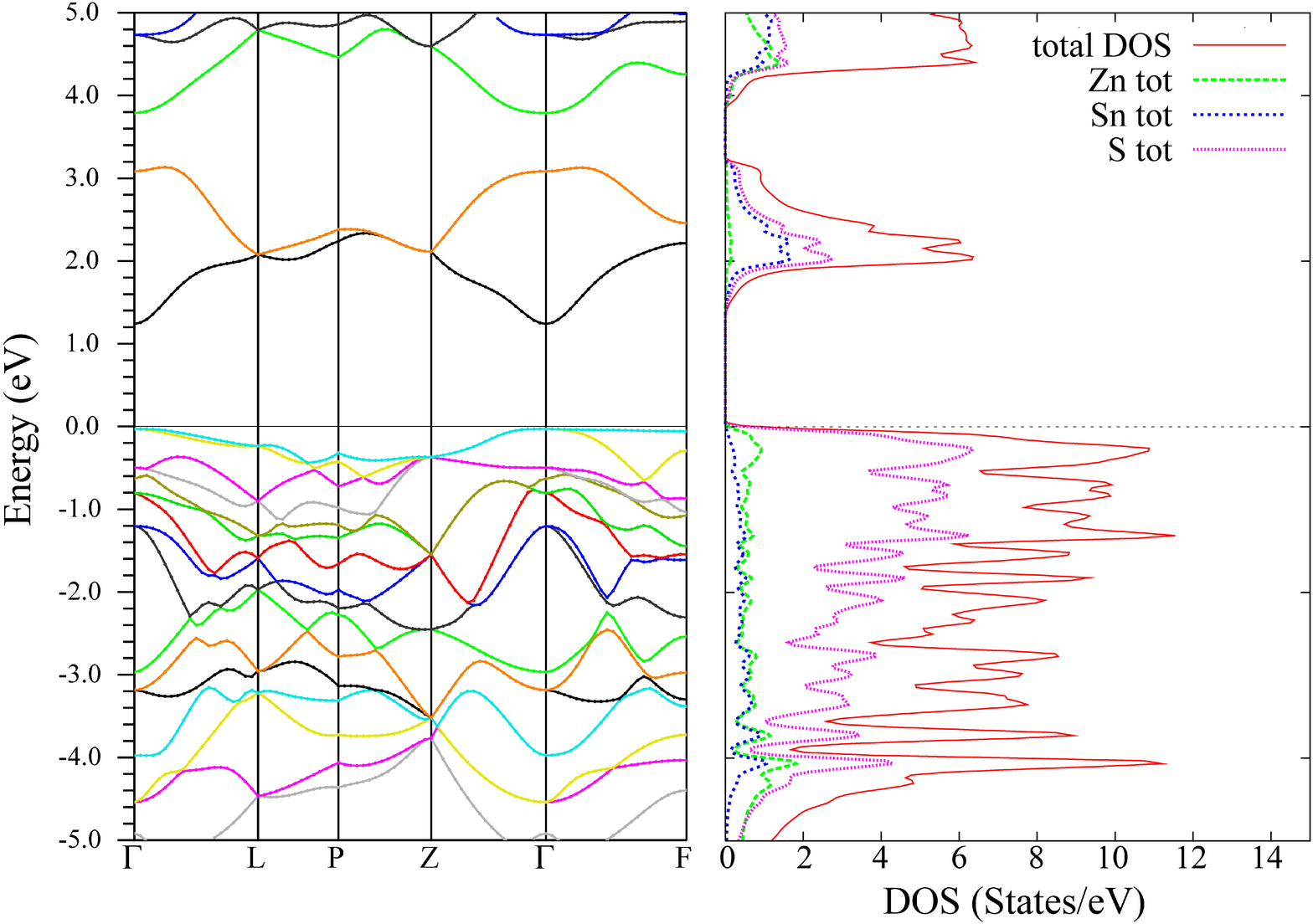}
   \caption{ Energy bands in the LN-phase (space group number 161) of ZnSnS$_3$,
     plotted along high symmetry directions, along with a plot of the density of states.
     The bands are calculated by using the mBJ potential for exchange and correlation.
     The zero of energy is taken to coincide with the valence band maximum.
  \label{fig:2}}
 \end{figure}

\section{Conclusions}\label{sec:4}

As anticipated, our calculations indicate that the substitution of sulfur for oxygen in 
ZnSnO$_3$ would reduce its band gap, particularly for the LN-phase. This reduction is 
sufficient to bring the absorption band of ZnSnS$_3$ into the visible to mid-infrared 
spectrum, thereby making it a suitable material for solar cell applications. 
An unexpected result of the substitution is that a  monoclinic phase 
becomes the most stable structure of ZnSnS$_3$. However, we assume that, 
as with ZnSnO$_3$, a suitable application of temperature and pressure may be used 
to transition ZnSnS$_3$ to its LN-phase. More work must be done to characterize 
the nature of this transition. The introduction of a stable monoclinic phase in 
ZnSnS$_3$ implies that one cannot simply assume that materials generated by substituting 
another element for oxygen in perovskite oxides will maintain the same relationships 
between phases. Future analysis should more thoroughly investigate the phases of 
ZnSnS$_3$, evaluating their stability at a variety of pressures. The transition pathway 
between phases should also be identified. Finally, the effect of the sulfur for oxygen 
substitution on other oxides such as ZnGeO$_3$ and ZnTiO$_3$ should be considered.

\acknowledgments

The authors gratefully acknowledge support by National Science Foundation
 under grant No. HRD-0932421 and NSF PREM Program: Cal State L.A. \& Penn
State Partnership for Materials Research and Education, award DMR-1523588.
We also acknowledge partial support by the Materials Simulation Center, 
at Penn-State and MRI facility.

\section*{Conflict of interest}

The authors report no conflict of interest in this research.



\begin{thebibliography}{12}

\bibitem{ref:Cao}
D. Cao et al., ``High-efficiency ferroelectric-film solar cells with an n-type Cu$_2$O cathode
buffer layer," \emph{Nano Lett.}, vol. 12, no. 6, pp. 2803--2809, 2012.

\bibitem{ref:Alexe}
M. Alexe and D. Hesse, ``Tip-enhanced photovoltaic effects in bismuth ferrite," 
\emph{Nature Commun.}, vol. 2, no. 1, Article ID 256, 2011.

\bibitem{ref:Qin}
M. Qin, K. Ao, and Y.C. Liang, ``High efficiency photovoltaics in nanoscaled ferroelectric thin films,"
\emph{Appl. Phys. Lett.}, vol. 93, no. 12, Article ID 122904, 2008.

\bibitem{ref:Choi1}
T. Choi, S. Lee. Y. Choi, V. Kiryukhlin, and S.-W. Cheong, ``Switchable ferroelectric diode and photovoltaic
effect in BiFeO$_3$," \emph{Science}, vol. 324, no. 5923, pp. 63--66 (2009).

\bibitem{ref:Butler}
K.T. Butler, J.M. Frost, and A. Walsh, ``Ferroelectric materials for solar energy conversion: photoferroics revisited,"
\emph{Energy Environ. Sci.}, vol. 8, no. 3, pp. 838--848, 2015.

\bibitem{ref:Yuan1}
Y. Yuan, Z. Xiao, B. Yang, and J. Huang,
``Arising applications of ferroelectric materials in photovoltaic devices,"
\emph{J. Mater. Chem. A}, vol. 2, no. 17, pp. 6027--6041, 2014.

\bibitem{ref:Yang}
S.Y. Yang et al.,
``Above-bandgap voltages from ferroelectric photovoltaic devices,"
\emph{Nature Nanotechnol}, vol. 5, no. 2, pp. 143--147, 2010.

\bibitem{ref:Inoue}
Y. Inoue, K. Sato, and H. Miyama,
``Photoassisted water decomposition by ferroelectric lead zirconate titanate ceramics with anomalous photovoltaic effects,"
\emph{J. Phys. Chem.}, vol. 90, no. 13, pp. 2809--2810, 1986.

\bibitem{ref:Young}
S.M. Young and A.M. Rappe,
``First-principles calculation of the shift current photovoltaic effect in ferroelectrics,"
\emph{Phys. Rev. Lett.}, vol. 109, no. 11, Article ID 116601, 2012.

\bibitem{ref:Huang}
H. Huang,
``Solar energy: Ferroelectric photovoltaics,"
\emph{Nature Photonics}, vol. 4, no. 3, pp. 134--135, 2010.

\bibitem{ref:Yuan2}
Y. Yuan et al.,
``Efficiency enhancement in organic solar cells with ferroelectric polymers,"
\emph{Nature Mater.}, vol. 10, no. 4, pp. 296--302, 2011.

\bibitem{ref:Bennett}
J.W. Bennett, I. Grinberg, and A.M. Rappe,
``New highly polar semiconductor ferroelectrics through d$^3$ cation-O vacancy
substitution in PbTiO$_3$: A theoretical study,"
\emph{J. Am. Chem. Soc.}, vol. 130, no. 51, pp. 17409--17412, 2008.

\bibitem{ref:Fridkin1}
V.M. Fridkin and B. Popov, ``Anomalous photovoltaic effect in ferroelectrics,"
\emph{Sov. Phys. Usp.}, vol. 21, no. 12, pp. 981--991, 1978.

\bibitem{ref:Fridkin2}
V.M. Fridkin, Book:
``Photoferroelectrics,"
\emph{Springer}, 1979.

\bibitem{ref:Glass}
A. Glass, D. Von der Linde, and T. Negran, ``High-voltage bulk photovoltaic effect and the
photoreactive process in LiNbO$_3$," \emph{Appl. Phys. Lett.}, vol. 25, no. 4, pp. 233--235, 1974.

\bibitem{ref:Chynoweth}
A.G. Chynoweth,
``Surface space-charge layers in barium titanate,"
\emph{Phys. Rev.}, vol. 102, pp. 705--714, 1956.

\bibitem{ref:Neumark}
G.F. Neumark,
``Theory of the Anomalous photovoltaic effect of ZnS,"
\emph{Phys. Rev.}, vol. 125, pp. 838--845, 1962.

\bibitem{ref:Choi2}
W.S. Choi, M.F. Chisholm, D.J. Singh, T. Choi, G.E. Jellison, and H.N. Lee,
``Wide band gap tunability in complex transition metal oxides by site-specific substitution,"
\emph{Nature Commun.}, vol. 3, Article number 689, 2012.

\bibitem{ref:Ehara}
S. Ehara et al.,
``Dielectric properties of Bi$_4$Ti$_3$O$_{12}$ below the Curie temperature,"
\emph{Jpn. J. Appl. Phys}, vol. 20, no. 5, pp. 877--881, 1981.

\bibitem{ref:Singh}
D.J. Singh, S.S.A. Seo, and H.N. Lee,
``Optical properties of ferroelectric Bi$_4$Ti$_3$O$_{12}$,"
\emph{Phys. Rev. B}, vol. 82, Article ID 180103, 2010.

\bibitem{ref:Jia}
C. Jia, Y.Chen, and W.F. Zhang,
``Optical properties of aluminum-, gallium-, and indium-doped Bi$_4$Ti$_3$O$_{12}$ thin films,"
\emph{J. Appl. Phys}, vol. 105, Article ID 113108, 2009.

\bibitem{ref:Arima}
T.Arima, Y. Tokura, and J.B. Torrance,
``Variation of optical gaps in perovskite-type 3d transition-metal oxides,"
\emph{Phys. Rev. B}, vol. 48, Article ID 17006, 1993.

\bibitem{ref:Grinberg}
I. Grinberg et al.,
``Perovskite oxides for visible-light-absorbing ferroelectric and photovoltaic materials."
\emph{Nature}, vol. 503, pp. 509--512, 2013.

\bibitem{ref:Basu}
S.R. Basu,
``Photoconductivity in BiFeO$_3$ thin films,"
\emph{Appl. Phys. Lett.}, vol. 92, Article ID 091905, 2008.

\bibitem{ref:Hauser}
A.J. Hauser et al.,
``Characterization of electronic structure and defect states of thin epitaxial BiFeO$_3$
films by UV-visible absorption and cathodoluminescence spectroscopies,"
\emph{Appl. Phys. Lett.}, vol. 92, Article ID 222901, 2008.

\bibitem{ref:Nechache1}
R. Nechache et al.,
``Growth, structure and properties of epitaxial thin films of first principles
predicted multiferroic Bi$_2$FeCrO$_6$,"
\emph{Appl. Phys. Lett.}, vol. 89, Article ID 102902, 2006.

\bibitem{ref:Nechache2}
R. Nechache, C. Harnagea, A. Pignolet, L.-P. Carignan, and D. M\'{e}nard,
``Epitaxial Bi$_2$FeCrO$_6$ multiferroic thin films,"
\emph{Phil. Mag. Lett.}, vol. 87, no. 3--4, pp. 231-240, 2007.

\bibitem{ref:Nechache3}
R. Nechache, C. Harnagea, S. Li, L. Cardenas, W. Huang, J. Chakrabartty, and F. Rosei,
``Bandgap tuning of multiferroic oxide solar cells,"
\emph{Nature Photonics}, vol. 9, pp. 61--67, 2015.

\bibitem{ref:Inaguma1}
Y. Inaguma, Y. Masashi, and T.Katsumata,
``A ploar oxide ZnSnO$_3$ with LiNbO$_3$ structure,"
\emph{J. Am. Chem. Soc.}, vol. 130, no. 21, pp. 6704--6705, 2008.

\bibitem{ref:Inaguma2}
Y. Inaguma, M. Yoshida, T. Tsuchiya, A. Aimi, K. Tanaka, T. Katsumata, and D. Mori,
``High-pressure synthesis of novel lithium niobate-type oxides,"
\emph{J. Phys.: Conf. Ser.}, vol. 215, no. 1, Article ID 012131, 2010.

\bibitem{ref:Inaguma3}
Y. Inaguma, K. Tanaka, T. Tsuchiya, D. Mori, T. Katsumata, T. Ohba, K. Hiraki, T.Takahashi, and H. Saitoh,
``Synthesis, structural transformation, thermal stability, valence state, and magnetic
and electronic properties of PbNiO$_3$ with perovskite and LiNbO$_3$-type structures,"
\emph{J. Am. Chem. Soc.}, vol. 133, no. 42, pp. 16920--16929, 2011.

\bibitem{ref:Arielly}
R. Arielly, W.M. Xu, E. Greenberg, G.K. Rozenberg, M.P. Pasternak, G. Garbarino,
S. Clark, and R. Jeanloz,
``Intriguing sequence of GaFeO$_3$ structures and electronic states to 70 GPa,"
\emph{Phys. Rev. B}, vol. 84, Article ID 094109, 2011.

\bibitem{ref:Shi}
Y. Shi, Y. Guo, X. Wang, A.J. Princep, D. Khalyavin, P. Manuel. Y. Michiue, A. Sato,
K.Tsuda, S. Yu, M. Arai, Y. Shirako, M. Akaogi, N. Wang, K. Yamamura, and A.T. Boothroyd,
``A ferroelectric-like transition in a metal,"
\emph{Nature Mater.}, vol. 12, pp. 1024--1027, 2013.

\bibitem{ref:Yusa}
H. Yusa, M. Akaogi, N. Sata, H. Kojitani, R. Yamamoto, and Y. Obishi,
``High-pressure transformations of ilmenite to perovskite, and lithium niobate
to perovskite in zinc germanate,"
\emph{Phys. Chem. Minerals}, vol. 33, no.3, pp. 217--226, 2006.

\bibitem{ref:Akaogi1}
M. Akaogi, H. Kojitani, H. Yusa, R. Yamamoto, M. Kido, and K. Koyama,
``High-pressure transitions and thermochemistry of MGeO$_3$ (M = Mg, Zn, and Sr) and\cite{ref:Borhade, ref:Mizoguchi}
Sr-silicates: systematics in enthalpies of formation of A$^{2+}$B$^{4+}$O$_3$ perovskites,"
\emph{Phys. Chem. Minerals}, vol. 32, no. 8, pp. 603--613, 2005.

\bibitem{ref:Zhang}
J. Zhang, B. Xu, Z. Qin, X.F. Li, and K.L. Yao,
``Ferroelectric and nonlinear optical properties of the LiNbO$_3$-type ZnGeO$_3$
from first-principles study,"
\emph{J. Alloys and Compounds}, vol. 514, pp. 113--119, 2012.

\bibitem{ref:Inaguma4}
Y. Inaguma, A. Aimi, Y. Shirako, D. Sakurai, D. Mori. H. Kojitani, M. Akaogi, and M. Nakayama,
``High-pressure synthesis, crystal structure, and phase stability relations of a LiNbO$_3$-type
polar titanate ZnTiO$_3$ and its reinforced polarity by the second-order Jahn-Teller effect,"
\emph{J. Am. Chem. Soc.}, vol. 136, no. 7, pp. 2748--2756, 2014.

\bibitem{ref:Syono1}
Y. Syono, S.-L. Akimoto, Y. Ishikawa, and Y. Endoh,
``A new high pressure phase of MnTiO$_3$ and its magnetic property,"
\emph{J. Phys. Chem. Solids}, vol. 30, no. 7, pp. 1665--1672, 1969.

\bibitem{ref:Ito}
E. Ito and Y. Matsui,
``High-pressure transformations in silicates, germanates and titanates with ABO$_3$ stoichiometry,"
\emph{Phys. Chem. Minerals}, vol. 4, no. 3, pp. 265--273, 1979.

\bibitem{ref:Ko}
J. Ko and C.T. Prewitt,
``High-pressure phase transition in MnTiO$_3$ from ilmenite to LiNbO$_3$ structure,"
\emph{Phys. Chem. Minerals}, vol 15, no. 4, pp. 355--362, 1988.

\bibitem{ref:Ross}
N.L. Ross, J. Ko, and C.T. Prewitt,
``A new phase transition in MnTiO$_3$: LiNbO$_3$-perovskite structure,"
\emph{Phys. Chem. Miner.}, vol. 16, no. 7, pp. 621--629, 1989.

\bibitem{ref:Syono2}
Y. Syono, H. Sawamoto, and S. Akimoto,
``Disordered ilmenite MnSnO$_3$ and its magnetic property,"
\emph{Solid State Commun.}, vol. 7, no. 9, pp. 713--716, 1969.

\bibitem{ref:Leinenweber1}
K. Leinenweber, W. Utsumi, Y. Tsuchida, T. Yagi, and K. Kurita,
``Unquenchable high-pressure perovskite polymorphs of MnSnO$_3$ and FeTiO$_3$,"
\emph{Phys. Chem. Miner.}, vol. 18, no. 4, pp. 244--250, 1991.

\bibitem{ref:Mehta}
A. Mehta, K. Leinenweber, A. Navrotsky, and M. Akaogi,
``Calorimetric study of high pressure polymorphism in FeTiO$_3$: The
stability of the perovskite phase,"
\emph{Phys. Chem. Miner.}, vol. 21, no. 4, pp. 207--212, 1994.

\bibitem{ref:Leinenweber2}
K. Leinenweber, J. Linton, A. Navrotsky, Y. fei, and J.B. Parise,
``High-pressure perovskites on the join CaTiO$_3$-FeTiO$_3$,"
\emph{Phys. Chem. Miner.}, vol. 22, no. 4, pp. 251--258, 1995.

\bibitem{ref:Hattori}
T. Hattori, T. Matsuda, T. Tsuchiya, T, Nagai, and T. Yamanaka,
``Clinopyroxene-perovskite phase transition of FeGeO$_3$ under high pressure and room temperature,"
\emph{Phys. Chem. Miner.}, vol. 26, no. 3, pp. 212--216, 1999.


\bibitem{ref:Leinenweber3}
K. Leinenweber, Y. Wang, T. Yagi, and H. Yusa,those
``An unquenchable perovskite phase of MgGeO$_3$ and comparison with MgSiO$_3$ perovskite,"
\emph{Am. Mineral.}, vol. 79, no. 1--2, pp. 197-199, 1994.

\bibitem{ref:Sleight}
A.W. Sleight and C.T. Prewitt,
``Preparation of CuNbO$_3$ and CuTaO$_3$ at high pressure,"
\emph{Mater. Res. Bull.}, vol. 5, no. 3, pp. 207--211, 1970.

\bibitem{ref:Akaogi2}
M. Akaogi, K. Abe, H. Yusa, H. Kojitani, D. Mori, and Y. Inaguma,
``High-pressure phase behaviors of ZnTiO$_3$: ilmenite-perovskite transition, decomposition of
perovskite into constituent oxides, and the perovskite-lithium niobate transition,"
\emph{Phys. Chem. Miner.}, vol. 42, no. 6, pp. 421--429, 2015.

\bibitem{ref:Borhade}
A.V. Borhade and Y.R. Baste,
``Study of photocatalytic asset of the ZnSnO$_3$ synthesized by green chemistry,"
\emph{Arabian Journal of Chemistry}, Online: http://dx.doi.org/10.1016/j.arabic.2012.10.001, 2012.

\bibitem{ref:Mizoguchi}
H. Mizoguchi and P.M. Woodward,
``Electronic structure studies of main group oxides possessing edge-sharing octahedra:
implications for the design of transparent conducting oxides,"
\emph{Chemistry of Materials}, vol. 16, no. 25, pp. 5233--5248, 2004.

\bibitem{ref:Kolb1}
B. Kolb and A. Kolpak,
``Zinc stannate as a solar cell material,"
\emph{Bull. Am. Phys. Soc.}, vol. 59, 2014.

\bibitem{ref:Kolb2}
 B. Kolb and A. Kolpak,
``First-principles design and analysis of an efficient, Pb-free ferroelectric photovoltaic
absorber derived from ZnSnO$_3$,"
\emph{Chem. Mater.}, vol. 27, no. 17, pp. 5899--5906, 2015.

\bibitem{ref:Kons}
C. Kons, A. Datta, and P. Mukherjee,
``Band-gap tuning in perovskite-type ferroelectric ZnSnO$_3$ by doping and
core shell approach for solar cell applications,"
\emph{Bull. Am. Phys. Soc.}, vol. 60, no. 1, 2015.

\bibitem{ref:Tran09}
F. Tran and P. Blaha, ``Accurate band gaps of semiconductors and insulators with a
semilocal exchange-correlation potential,"
\emph{Phys. Rev. Lett.}, vol. 102, no. 22, Article ID 226401, 2009.

\bibitem{ref:Oganov1}
A.R. Oganov and C.W. Glass,
``Crystal structure prediction using ab initio evolutionary techniques: Principles and applications,"
\emph{J. Chem. Phys.}, vol. 124, no. 24, Article ID 244704, 2006.

\bibitem{ref:Lyakhov}
A.O. Lyakhov, A.R. Oganov, H.T. Stokes, and Q. Zhu,
``New developments in evolutionary structure prediction algorithm USPEX,"
\emph{Comp. Phys. Commun.}, vol. 184, no. 9, pp. 1172--1182, 2013.

\bibitem{ref:Oganov2}
A.R. Oganov, A.O. Lyakhov, and M. Valle,
``How evolutionary crystal structure prediction works--and why,"
\emph{Acc. Chem. Res.}, vol. 44, no. 3, pp. 227--237, 2011.

\bibitem{ref:calypso}
Y. Wang, J. Lv, L. Zhu, and Y. Ma,
``CALYPSO: A method for crystal structure prediction,"
\emph{Comp. Phys. Commun.}, vol. 183, no. 10, pp. 2063--2070, 2012.

\bibitem{ref:Kreese1}
G. Kresse and J. Hafner, ``Abinitio Molecular-Dynamics for Liquid-Metals,''
\emph{Phys. Rev. B}, vol. 47, no. 1, pp. 558--561, 1993.

\bibitem{ref:Kreese2}
G. Kresse and J. Furthmuller, ``Efficiency of ab-initio total energy calculations for 
metals and semiconductors using a plane-wave basis set,''
\emph{Computational Mat. Sci.}, vol. 6, no. 1, pp. 15-50, 1996.

\bibitem{ref:Kreese3}
G. Kresse and D. Joubert, ``From ultrasoft pseudopotentials to the projector augmented-wave method,''
\emph{Phys. Rev. B}, vol. 59, no. 3, pp. 1758-1775, 1999.

\bibitem{ref:Blaha}
P. Blaha, K. Schwarz, G. Madsen, D. Kvasnicka, and J. Luitz, User's Guide:
``WIEN2K: an augmented plane wave + local orbitals program for
calculating crystal properties," 2001.

\bibitem{ref:Monkhorst}
 H. Monkhorst and J. Pack, ``Special points for Brillouin-zone integrations,"
\emph{Phys. Rev. B}, vol. 13, no. 12, pp. 5188--5192, 1976.

\bibitem{ref:Brehm}
J.A. Brehm, J.W. Bennett, M.R. Schoenberg, I. Grinberg, and A.M. Rappe,
''The structural diversity of ABS$_3$ with d$^0$ electronic configuration
for the B cation,`` \emph{J. Chem. Phys.}, vol. 140, no. 22, Article ID 224703, 2014.

\bibitem{ref:Resta}
R. Resta, ``Macroscopic electric polarization as a geometric quantum phase,"
\emph{Europhys. Lett.}, vol. 22, no. 2, pp. 133--138, 1993.

\bibitem{ref:King}
R.D. King-Smith and D. Vanderbilt, ``Theory of polarization of crystalline solids,"
\emph{Phys. Rev. B}, vol. 47, no. 3, pp. 1651--1654, 1993.






















\end{thebibliography}
\end{document}